\documentclass[9pt,twocolumn,twoside]{pnas-new}

\templatetype{pnasresearcharticle} 

\title{Spatial heterogeneities shape collective behavior of signaling amoeboid cells }

\author[a,1]{T. Eckstein}
\author[a,1]{E. Vidal-Henriquez}
\author[a]{A. Bae}
\author[a,2]{A. Gholami}

\usepackage{soul}

\affil[a]{Max Planck Institute for Dynamics and Self-Organization, Am Fassberg 17, 37077 G\"{o}ttingen, Germany} 

\leadauthor{Gholami} 

\significancestatement{Large scale cAMP waves synchronize the population of starving Dictystelium discoideum cells and enable them to aggregate and form a multi-cellular organism. In nature, Dictyostelium cells are exposed to external obstacles which can significantly influence the wave generation process. Here, we confirm experimentally that spatial heterogeneities indeed shape the collective behavior of Dictyostelium cells and used a reaction-diffusion model to identify the underlying mechanism for unexcitable obstacles to act as a wave source. We expect our investigation to be generic to other excitable media governed by reaction-diffusion equations, where spatial heterogeneities can play a critical role in spatio-temporal dynamics of non-linear waves.
}
\authorcontributions{T.E. performed experiments, E.V.H. designed and performed numerical simulations, T.E E.V.H., A.B and A.G analyzed data, A.B and A.G designed research, T.E, E.V.H., A.B., and A.G wrote the paper.}
\authordeclaration{The authors declare no conflict of interest.}
\equalauthors{\textsuperscript{1}T.E. and E.V.H.contributed equally to this work.}
\correspondingauthor{\textsuperscript{2}To whom correspondence should be addressed. E-mail: azam.gholami@ds.mpg.de }


\begin{abstract}
We present novel experimental results on pattern formation of signaling Dictyostelium discoideum amoeba in the presence of a periodic array of millimeter-sized pillars. We observe concentric cAMP waves that initiate almost synchronously at the pillars and propagate outwards. These waves have higher frequency than the other firing centers and dominate the system dynamics. The cells respond chemotactically to these circular waves and stream towards the pillars, forming periodic Voronoi domains that reflect the periodicity of the underlying lattice. We performed comprehensive numerical simulations of a reaction-diffusion model to study the characteristics of the boundary conditions given by the obstacles. Our simulations show that, the obstacles can act as the wave source depending on the imposed boundary condition. Interestingly, a critical minimum accumulation of cAMP around the obstacles is needed for the pillars to act as the wave source. This critical value is lower at smaller production rates of the intracellular cAMP which can be controlled in our experiments using caffeine. Experiments and simulations also show that in the presence of caffeine the number of firing centers is reduced which is crucial in our system for circular waves emitted from the pillars to successfully take over the dynamics. These results are crucial to understand the signaling mechanism of Dictyostelium cells that experience spatial heterogeneities in its natural habitat. 
\end{abstract}

\dates{This manuscript was compiled on \today}
\doi{\url{www.pnas.org/cgi/doi/10.1073/pnas.XXXXXXXXXX}}

\begin{document}

\verticaladjustment{-2pt}

\maketitle
\thispagestyle{firststyle}
\ifthenelse{\boolean{shortarticle}}{\ifthenelse{\boolean{singlecolumn}}{\abscontentformatted}{\abscontent}}{}


\dropcap{A} fundamental process occurring in reaction-diffusion excitable systems is the propagation of nonlinear waves~\cite{holden2013nonlinear,grindrod1991patterns,meron1992pattern,zykov1987simulation}. Examples of
such waves include chemical waves in the Belousov-Zhabotinsky reaction~\cite{zaikin1970concentration,manz2002excitation}, waves of CO oxidation on Pt catalytic surfaces~\cite{imbihl1995oscillatory}, electrical waves in retinal and cortical nerve tissue~\cite{gorelova1983spiral}, waves in heart muscle~\cite{davidenko1992stationary,allessie1973circus}, and cAMP (cyclic adenosine monophosphate) waves in starved population of {\it Dictyostelium discoideum} ({\it D.d.}) amobae~\cite{gerisch1965stadienspezifische,devreotes1983quantitative,laub1998molecular}. One of the important questions concerning nonlinear
waves is how they propagate in the presence of obstacles.  Various aspects of this question, such as the influence of one or few large obstacles or large number of small obstacles  on
wave propagation in excitable media, have been extensively studied over the years~\cite{agladze1994rotating,panfilov1993effects,ito1991spiral,ten2005wave,bar2002pattern,steinbock1992chemical,luther2011low}.  

Here, we report experimental and numerical results on the spatial-temporal dynamics of population of {\it D.d.} cells in the presence of non-excitable obstacles. This organism, naturally occurring in forest soil, is an important model system for the study of chemotaxis, cell differentiation, and morphogenesis~\cite{chisholm2004insights}. Starvation of {\it D.d.} cells induces a developmental program in which cells align to form head-to-tail streams by signaling to each other with cAMP. Cells initiate the process by sending out pulsatile signals with a periodicity of several minutes, which propagate as waves. With time circular and spiral patterns form. 
Cells respond chemotactically to cAMP waves that guide cell movement towards the signaling centers and form multi-cellular centimeter-scale Voronoi domains. The corresponding wave sources in each domain then act as aggregation centers, which eventually transform into millimeter long slugs and finally into fruiting bodies bearing spores for long-term survival and long-range dispersal~\cite{saran2002camp,laub1998molecular}. In their natural habitat, populations of starving cells are exposed to spatial  heterogeneities that will profoundly influence the processes of wave generation and propagation.  In nature the obstacles are randomly distributed in 3D, but  as a first step towards understanding, we look at a simpler system of cells in a 2D geometry with a periodic arrangement of obstacles.
\begin{figure}
	\centering
	\includegraphics[width=0.9\columnwidth]{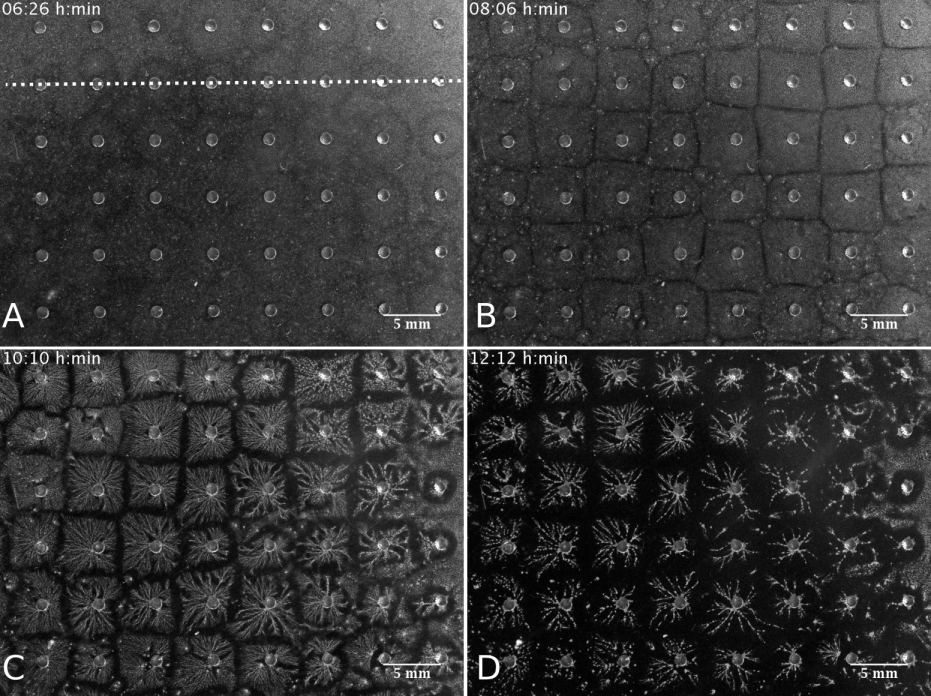}
	\caption{Top view of {\it Dictyostelium} cells on a macro-pillar array. A) Concentric waves initiate around the pillars and propagate outwards. (B-D) The amoebas respond chemotactically to the circular waves and stream towards the posts. This leads to the formation of regular Voronoi domains around the pillars. Timestamps denote time since start of starvation.}
	\label{Fig:Quadratic}
\end{figure}
\section*{Experimental Results}
Our quasi 2D geometry consists of a regular array of millimeter-sized pillars that control spatio-temporal dynamics of a population of uniformly distributed {\it D.d.} cells ({\it Material and Methods}). As shown in Fig.~\ref{Fig:Quadratic}, these spatial heterogeneities induce circular waves centered on the posts that trigger chemotactic cell movement towards the pillars. This leads to the formation of periodic Voronoi domains that reflect the periodicity of the underlying macro-pillar array. Interestingly, we observed synchronized circular waves and regular Voronoi domains only in the presence of caffeine. Moreover, a similar phenomenon was observed in the experiments where we used holes instead of pillars as obstacles.
\begin{figure}
	\centering
	\includegraphics[width=0.9\columnwidth]{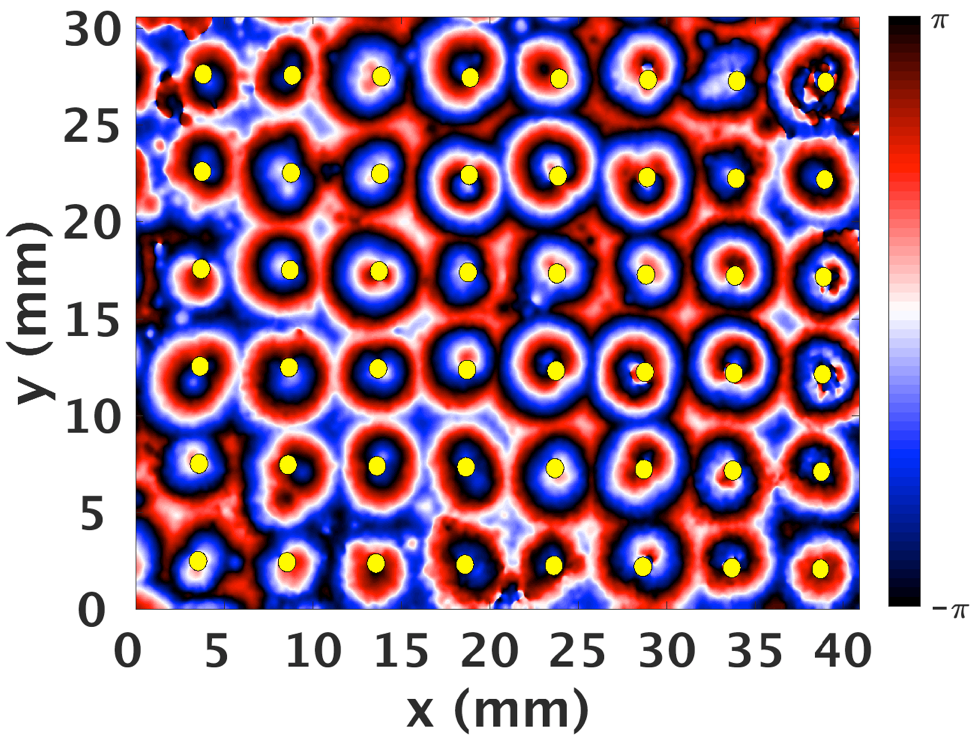}
	\caption{Phase map of the observed spatial-temporal pattern showing the formation of circular waves around the pillars. See supplementary movie 4 for the time evolution of the phase map.}
	\label{i:rect_phase}
\end{figure}
\begin{figure}[t]
	\centering
	\includegraphics[width=0.9\columnwidth]{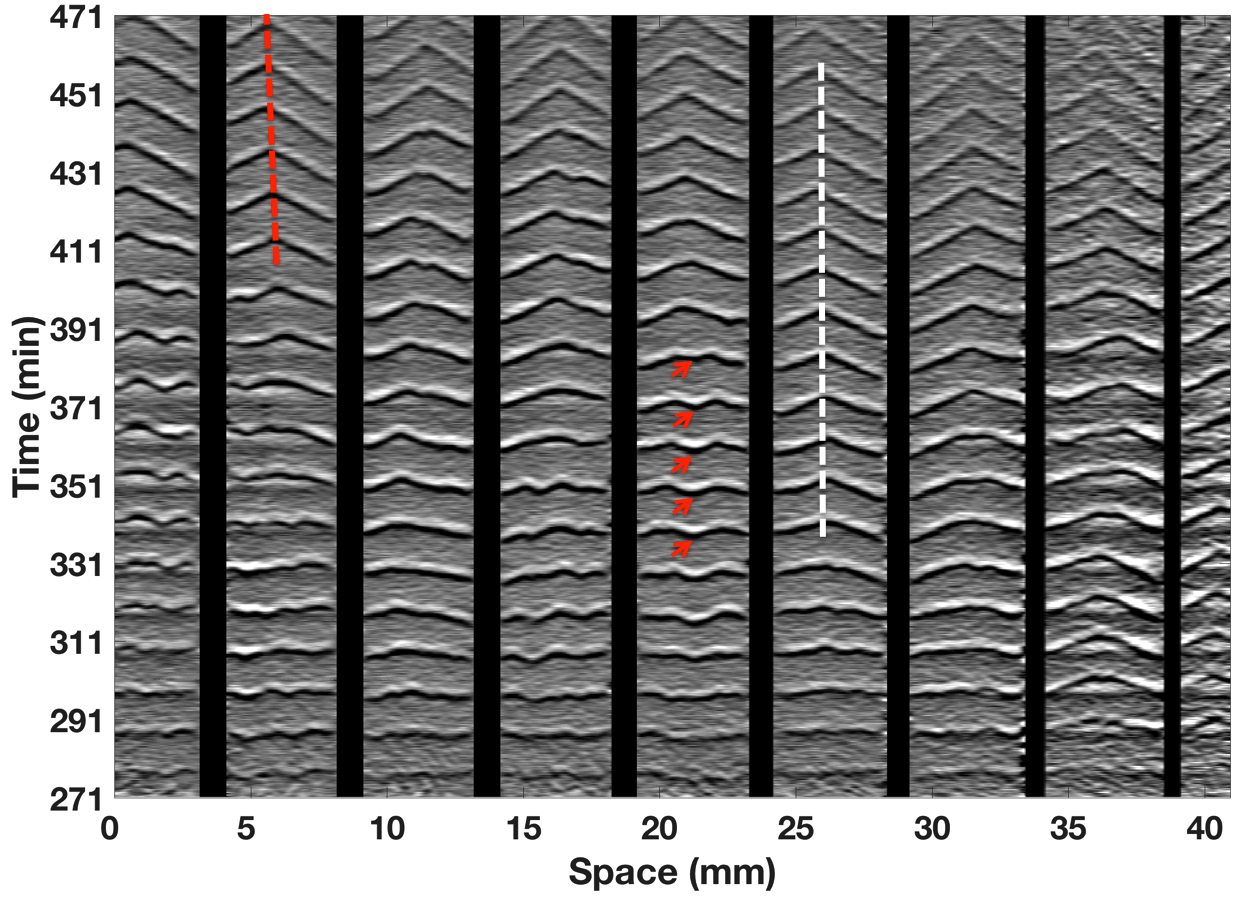}
	\caption{Space-time plot of the experiment in Fig.~\ref{Fig:Quadratic}. The light intensity from supplemental movie 3 is stacked up along the white dashed line shown in Fig.~\ref{Fig:Quadratic}(a). The black bars show the position of the pillars. The red arrows show a firing center other than the pillars that has a lower frequency and disappears with time. The white dashed line traces the annihilation points of two waves that initiate on the neighboring post and have a similar frequency but a phase shift. The red dashed line shows the slow drift of the annihilation point towards the post with smaller frequency (left pillar).}
	\label{Fig:Kymo}
\end{figure}

In the absence of pillars, spiral or target patterns emerge at random locations on the PDMS substrate (see supplemental movie 1). This is in contrast with patterns in the presence of macro-pillars, where waves originate  at the posts and propagate outwards, as shown in Fig.~\ref{Fig:Quadratic}. The four successive snapshots reveal circular waves centered around the pillars, cell streaming towards the posts, formation of regular Voronoi domains, and cell aggregation, respectively. In our experiments, we observed that concentric waves develop almost synchronously around the pillars and since they have a slightly higher frequency, dominate over the other firing centers. Waves propagate outwards from the posts and trigger chemotactic movement of the cells towards the pillars. As a result, periodic Voronoi domains form around the pillars.

The concentric waves around the pillars are well visible in the corresponding spatial phase map in Fig.~\ref{i:rect_phase}. The phase map shows that (i) the circular waves are slightly off-center from the pillars (ii) the territories that each circular wave propagates before annihilation have different sizes, and (iii) the frequency and phase of the circular waves can vary between the pillars. 
\begin{figure}
	\centering
	\includegraphics[width=0.9\columnwidth]{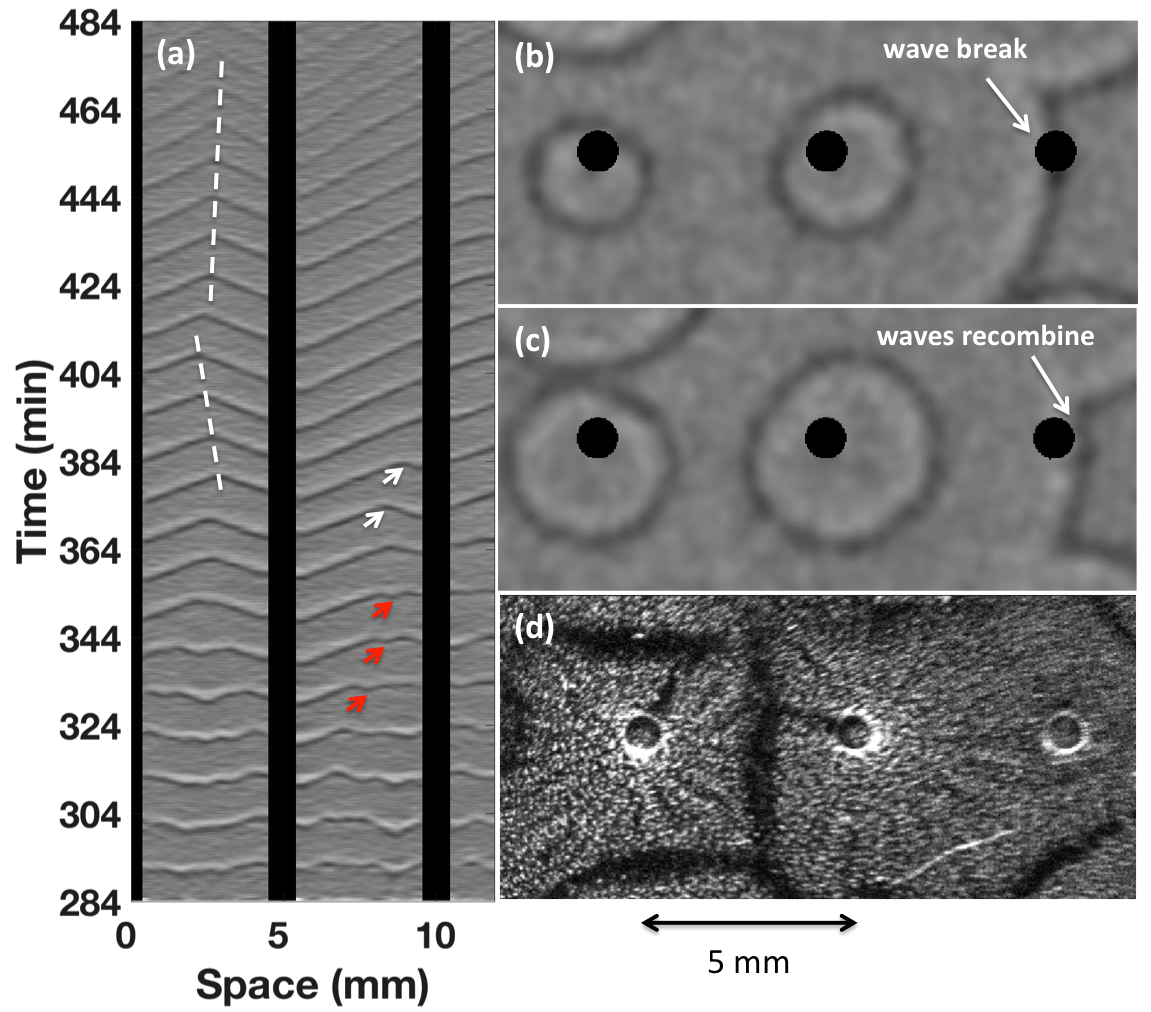}
	\caption{A) Space-time plot in an experiment with pillars of 50 $\mu$m high (supplemental movie 5). Only part of the kymograph with 3 pillars is shown. The red arrows point to a firing center other than the pillars which has a lower frequency and disappears with time. The white dashed lines show the movement of the annihilation points towards the pillars with lower wave frequency. The white arrows show the collision of the annihilation point with the right pillar. This pillar fails to emit its own circular cAMP wave and only breaks the wave front initiated from the neighboring pillar (B-D) As the wave front collides with the pillar on the right, it breaks and recombine again. As a result no boundary between two neighboring pillars form.}
	\label{Fig:Kymo-3pillars}
\end{figure}
%
%

Figure~\ref{Fig:Kymo} shows the space-time plot along the white dashed line in Fig.~\ref{Fig:Quadratic}A. For this kymograph, the processed images in supplemental movie 3 are used. First, we observe synchronized bulk oscillations (with a period around 10 min). We attribute this to initial starvation of the cells in a shaking suspension that leads to cell synchronization. After about 60 min, waves are initiated almost synchronously from the pillars and propagate outwards. They annihilate as they collide with each other. These annihilation areas define the boundaries of the regular Voronoi domains. In an ideal experiment, where the circular waves have the same frequency and phase, the size of the quadratic Voronoi domains is the same as the pillar spacing ($\sim$5 mm). However, in general there is a frequency and phase difference between concentric waves originating from neighboring posts. If two neighboring waves have almost the same frequency with a phase shift, due to this phase difference, the annihilation point of two neighboring circular waves is not located exactly at the middle of two pillars, but are rather located closer to the pillar with a phase delay. This is shown by the white dashed line in Fig.~\ref{Fig:Kymo}.  Moreover, often in the experiments, there is also a small frequency difference between the waves initiating at the neighboring pillars. Consequently, the annihilation point of two neighboring concentric waves shifts towards the pillar with smaller frequency. This event is shown by the red dashed line in Fig.~\ref{Fig:Kymo} and white dashed lines in Fig.~\ref{Fig:Kymo-3pillars}A. The drift velocity of the annihilation point can be calculated to be $v_w (f_2-f_1)/(f_2+f_1)$, where $v_w$ is the wave propagation velocity and $f_1$ and $f_2$ are the wave frequencies at the two neighboring posts ({\it see supplemental information}). In our experiments, $v_w$ is of the order of 0.4 mm/min and the frequencies $f_1$ and $f_2$ are roughly 1/10 min$^{-1}$ and 1/9 min$^{-1}$, which gives $v_\text{drift}$ to be of the order of 0.02 mm/min. If the frequency difference persists, eventually the wave front collides with the pillar and breaks as is shown by white arrows in Figs.~\ref{Fig:Kymo-3pillars}A, B. They recombine after passing through the obstacle and propagate further (see Fig.~\ref{Fig:Kymo-3pillars}C). In this case, the obstacle only breaks the propagating wave front and there is no boundary formed between the middle and the right pillars in Fig.~\ref{Fig:Kymo-3pillars}D. 

\begin{figure}[htbp]
	\centering
    \includegraphics[width=0.5\textwidth]{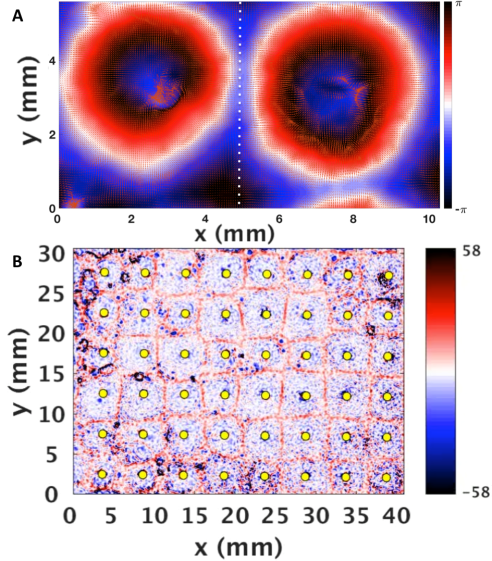}
	\caption{A) Gradient vectors of the phase map around two neighboring posts. The white dashed line shows the location where two emitted waves collide and the vector field changes direction. The vectors are scaled 50 times larger than the original values for better visibility. B) Laplacian of the phase map in unit of $1/mm^2$. The gradient vector field switches direction at the boundary of the Voronoi domains, thus Laplacian  has higher values at these boundaries.}
		\label{Fig:VectorField}
\end{figure}

In Fig.~\ref{Fig:VectorField}A, we calculated the gradient vectors of the phase map around two neighboring posts. The vector field changes direction where the concentric waves meet and annihilate each other. Thus, the Laplacian of the phase map $\phi$, defined as $\partial^2\phi(x,y)/\partial x^2+\partial^2\phi(x,y)/\partial y^2$, gets extreme values at the collision regions of two neighboring emitted waves and defines the boundaries of the Voronoi domains (see Fig.~\ref{Fig:VectorField}B).
%
%

\begin{figure}[htbp]
	\centering
	\includegraphics[width=0.4\textwidth]{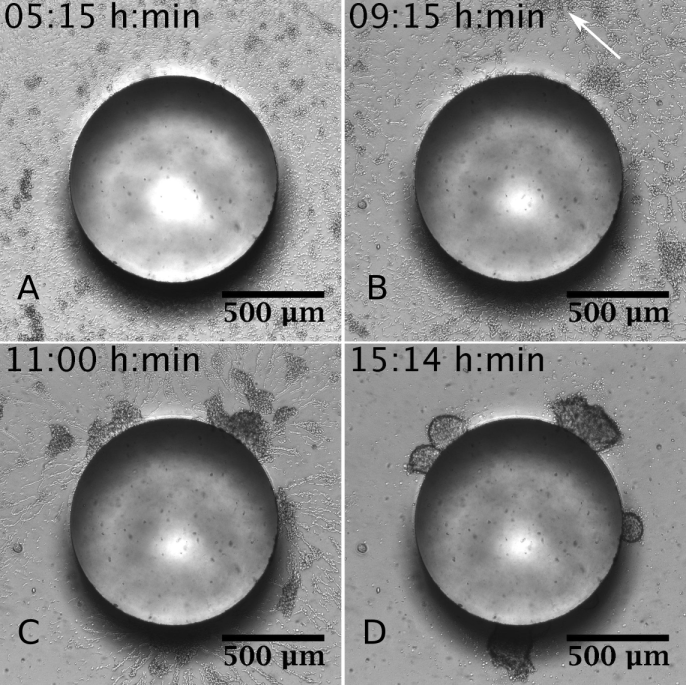}
		\caption{A closer look at the pillars with a bright field microscope. A) Initial distribution of the cells show small clumps. B-D) Cells chemotactically move towards the center of the concentric waves that are sometimes slightly off-center from the pillars. During the streaming process, the cells join small clusters to make larger ones and eventually large clusters aggregate on the post (see supplemental movie 6). Timestamps denote time since start of starvation.}
		\label{i:deltavision}
\end{figure}

Next, we used bright field microscopy to closely look at the wave propagation and cell streaming in the vicinity of the posts (see supplemental movie 6). A higher cell density around the pillars after plating the cells could explain the higher wave frequency of the waves initiating from the posts. Our extensive bright field observations didn't confirm a significant cell accumulation around the pillars. However, based on our numerical simulations, in the presence of caffeine even a tiny cAMP accumulation in the vicinity of the obstacles is enough to trigger formation of concentric waves around the posts. Possible cell attachment to the side walls of the pillars which is considered in simulations with a small constant value of cAMP on the obstacles, is a plausible mechanism for the formation of circular waves around the posts (see section Numerical Simulations). Moreover, we emphasize that initial starvation of the cells for 4 hours in a shaking suspension, leads to the formation of very small clusters that are visible in Fig.~\ref{i:deltavision}A. As we mentioned before, concentric waves around the pillars are not always all centered at the pillars, but are slightly off-center. Therefore, the cells first stream towards a point in the vicinity of the pillar (arrow in Fig.~\ref{i:deltavision}B) and then aggregate at the pillar itself. 

\begin{figure}
	\centering
	\includegraphics[width=0.4\textwidth]{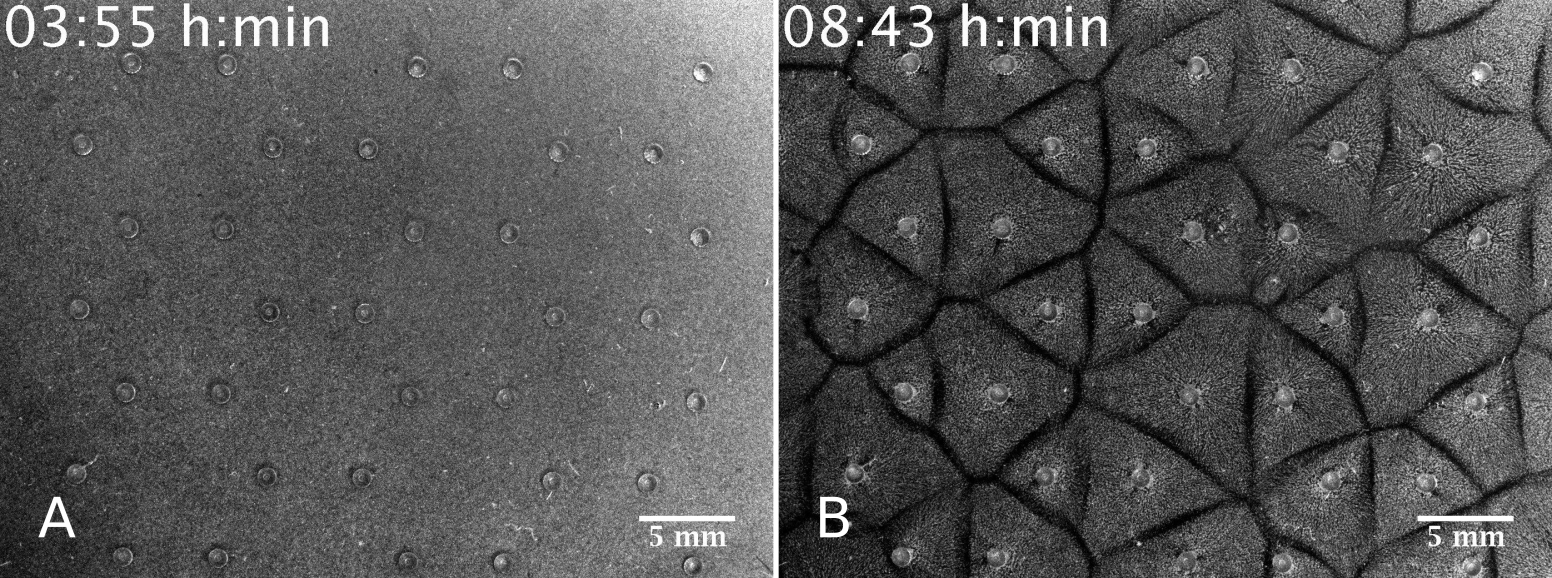}
	\caption{Different periodic arrangements of the pillars. Hexagonal and triangular Voronoi domains formed around the pillars with A) triangular and B) hexagonal arrangement, respectively (see supplemental movie 9 and 10).}
	\label{i:hex_tria}
\end{figure}

The phenomenon of initiation of synchronized circular waves around the pillars and formation of regular Voronoi domains were robust with respect to the arrangement of the pillars. We repeated our experiments with the same pillar size and spacing, but triangular and hexagonal arrangement of the posts, and observed hexagonal and triangular Voronoi domains, respectively (see Fig.~\ref{i:hex_tria}). 
\section*{Numerical Simulations}
\begin{figure}[htbp]
	\centering
	\includegraphics[width=\columnwidth]{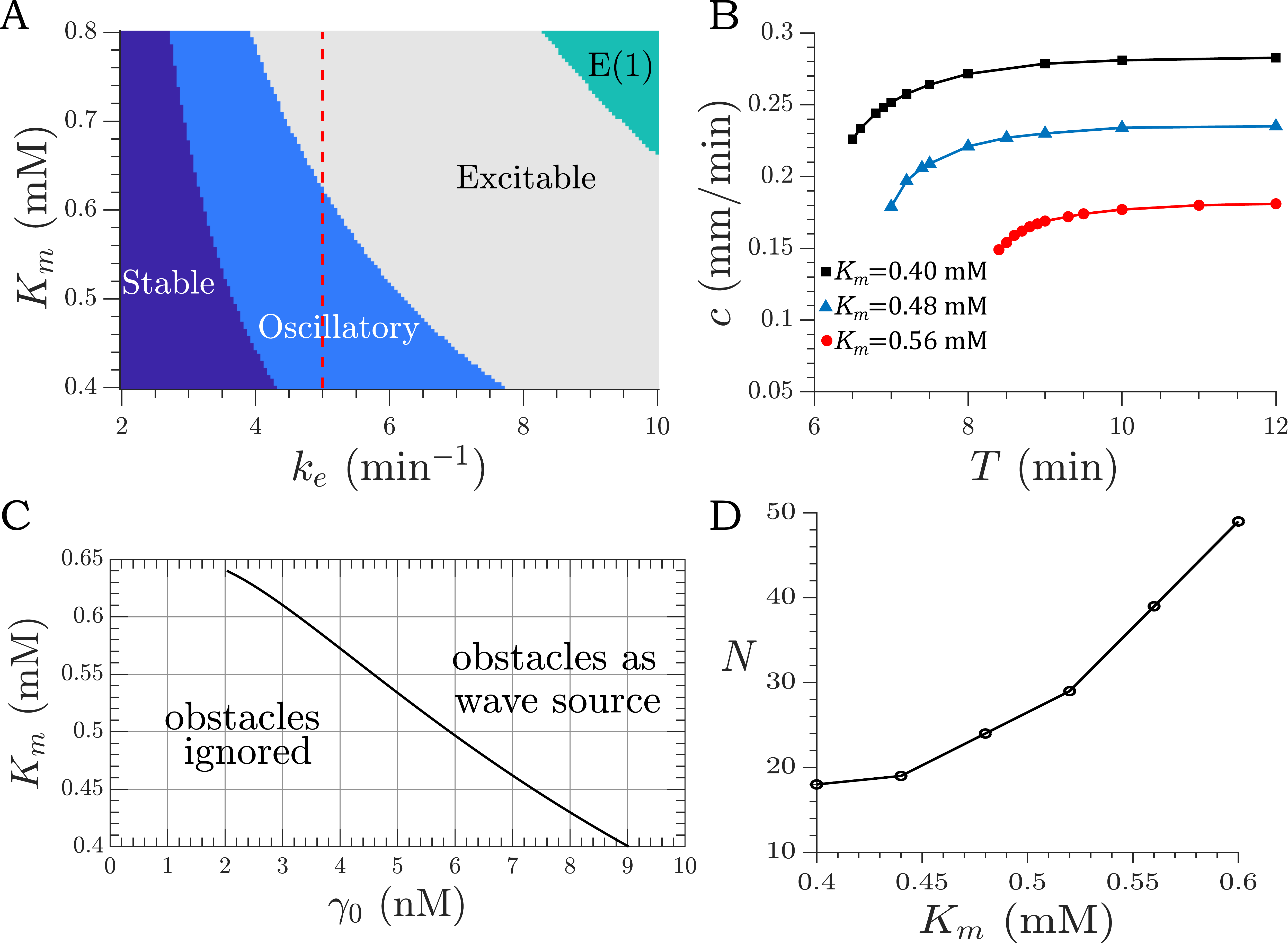}
	\caption{A) Phase diagram of the reaction-diffusion model used for simulating the experimental setup. On the area marked as Stable, the system has one solution, which is stable. In the Oscillatory region, the system shows one unstable steady state surrounded by a limit circle. In the Excitable regime, the system has 3 steady states, two unstable ones and a stable one, which is excitable. In the regime marked as E(1) the system shows one steady state, which is excitable. Stability calculated through linear analysis, and excitability through no-space simulations. Red dashed line shows the path that $K_m$ is changed in parts B, C and D at fixed value of $k_e=5.0$  min\textsuperscript{-1}. B) Dispersion relations of the supported wavetrains for 40\% surface coverage, where $T$ is wave period and $c$ wave velocity. Waves with periods below those shown, do not get relayed by the system. C) Effect of a boundary condition $\gamma_0$: for higher values of $K_m$ the minimum amount of $\gamma_0$ required to produce a target center decreases. D) Minimum number $N$ of consecutive cells (in 1-D) needed to produce a cluster with self-sustained oscillations.}
	\label{Fig:DispersionRelation}
\end{figure}

We performed numerical simulation of the model proposed by Martiel and Goldbeter for the creation and relay of cAMP in {\it D.d.} \cite{martiel1987model,tyson1989spiral}. Many approaches have been used to create spirals and target centers in this model, most of which define the position of the localized structures through perturbations or diversity of developmental stage among the cells  \cite{lauzeral1997desynchronization,geberth2009predicting}. To avoid any pre-location of the structure centers we used a dynamical approach for center creation ({\it see Materials and Methods}) where the cell locations are discretized inside grids, thus a grid containing a cell is an occupied one and can produce and relay cAMP, while the empty grids without cells can only degrade the signal via external phosphodisterase. This mixture of occupied and unoccupied areas on the system breaks the system homogeneity and allows clusters of local higher cell density to become target centers. The lower density areas are still capable of sustaining waves, thus the waves generated by these clusters get relayed by the rest of the system. We measured the dispersion relation of such waves and showed that they have the behavior of trigger waves \cite{aliev1994dynamics}(see Fig. \ref{Fig:DispersionRelation}B). These dispersion relations at different cell densities also showed that the wave velocity increases with cell density, which has been reported~\cite{van1996spatial} to be necessary to produce aggregation streams. 

\begin{figure}[htbp]
	\centering
	\includegraphics[width=0.9\columnwidth]{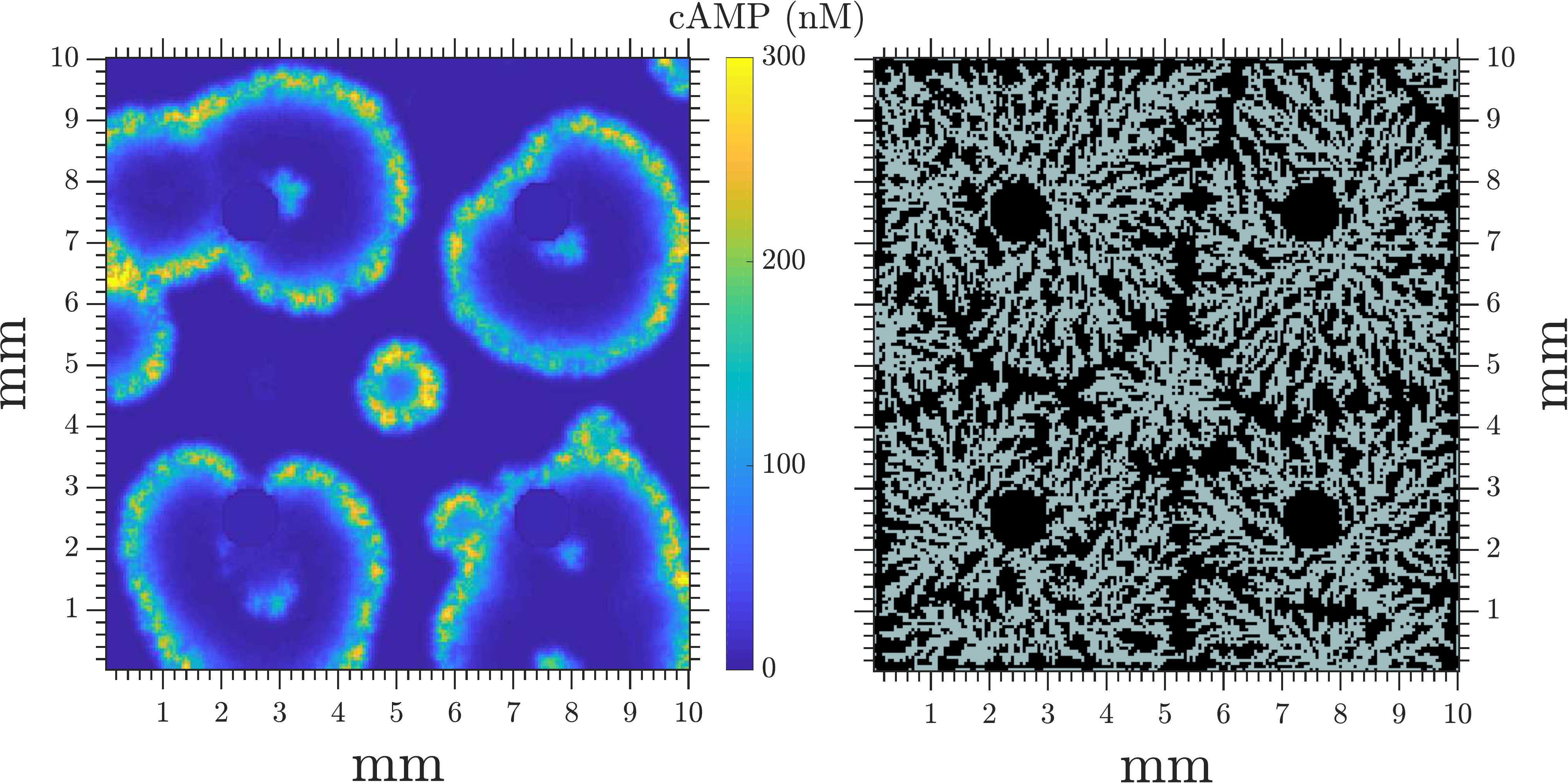}
	\caption{Numerical simulations of cAMP waves with four pillars as obstacles with fixed boundary condition $\gamma_0=6$ nM, $K_m=0.504$ mM, $k_e=6.0$ min\textsuperscript{-1}, $70\%$ of cell surface coverage.  
	 A) Concentric waves of cAMP coming out of the pillars. B) Cell distribution after $t=100$ min, showing Voronoid domains around the pillars. Grey squares show grid points containing a cell, black squares show the empty ones.}
	\label{Fig:StreamingSimulations}
\end{figure}

We observed in our simulations that a no flux boundary condition at the obstacles is not enough to produce centers. Similarly, when this boundary condition is applied, spirals only break after colliding with the obstacles and recombine again (supplemental movie 20). Other boundary condition that has been shown to produce traveling waves in numerical simulations is the Dirichlet boundary condition~\cite{sherratt2003periodic}, where the boundary is held to a fixed value. In our system, a fixed value of cAMP at the obstacles creates wave trains emitted from the obstacles if the fixed value is larger than a threshold, as it is shown in Fig. \ref{Fig:StreamingSimulations} and the supplemental movie 30. This minimum amount of cAMP needed for the obstacles to act as a wave source depends on the system parameters. We also observed that slightly higher density of amoebas around the pillars can trigger the generation of wave centers at the obstacles (supplemental movie 40).

To account for the effects of caffeine used in our experiments (see Materials and Methods and Discussion for the amounts and effects of caffeine, respectively), we varied the parameter $K_m$ which is the Michaelis constant of the reaction in which ATP produces intracellular cAMP. Increasing this parameter reduces the affinity between ATP and the enzyme adenylate cyclase, thus reducing the production rate of intracellular cAMP. We performed linear stability analysis of the Martiel-Goldbeter model with modifications in $K_m$ and characterized its different regimes, which are shown in Fig. \ref{Fig:DispersionRelation}A, the other parameter we varied is the degradation rate of external phosphodiesterase $k_e$. We chose the parameters such that the system is in the oscillatory regime, i.e. a stable limit cycle exists, and the cell coverage (ratio of occupied grids to total number of grids) is high enough for the waves to get relayed. At higher values of $K_m$, which is equivalent to higher concentration of caffeine in our experiments, the number of firing centers in the system is decreased (supplemental movie 50). The reason is that the minimum cluster size (measured as consecutive cells in a 1D setting) required to produce a self-sustained oscillatory center is increased (see Fig.~\ref{Fig:DispersionRelation}D). Increasing the amount of caffeine (higher $K_m$ values) also decreased the wave velocity and increased the minimum wave period the system can sustain. Different dispersion relations showing this effect are depicted in Fig. \ref{Fig:DispersionRelation}B. 
We studied the boundary effect in this system and found that at higher caffeine concentration the trigger waves require a smaller amount of cAMP to be triggered, thus the obstacles act as a wave source at smaller values of the fixed boundary $\gamma_0$ (see Fig. \ref{Fig:DispersionRelation}C).  
\section*{Discussion}
The results presented here show that external obstacles can significantly influence the wave generation process in starving populations of {\it D.d.} cells. We observed circular waves that initiate almost synchronously at the pillars and propagate outwards. Chemotactically competent cells detect the cAMP gradient and crawl towards the posts, forming a periodic array of Voronoi domains. This phenomenon is also observed for triangular and hexagonal arrangement of the pillars leading to the formation of hexagonal and triangular Voronoi domains, respectively. We also changed the material of the pillars from PDMS to Plexiglas (PMMA) and again observed the regular streaming domains. Most of our experiments are performed with pillars of 1 mm diameter, but we observed a similar phenomenon with pillars of 1.5 mm diameter (supplemental movie 7), shallow holes (100 $\mu m$) in PDMS (supplemental movie 8), and smaller center-to-center spacing of the pillars (3.75 mm instead of 5 mm). The phenomenon is also robust with respect to the pillar height since we observed regular Voronoi domains with pillar's height down to 50 micron (see Fig.\ref{Fig:Kymo-3pillars} and supplemental movie 5). The concentric waves emitted from pillars have a slightly higher frequency than the other firing centers, thereby dominating the system dynamics. Cells attached to the side walls of the PDMS can trigger higher frequency waves. In our simulations, we included this effect by assuming either a fixed value of cAMP around the posts or slighly higher cell accumulation in the vicinity of pillars (see supplemental Movie 500). Another scenario which requires extra bio-chemical analysis, is the possibility of accumulation or depletion of any chemical (such as phosphodiestrase) in the vicinity of the pillars. Our numerical simulations show that in the case of phosphodiestrase depletion, pillars can act as wave source in the system that might drive oscillations in the vicinity of the obstacles~\cite{stich2002complex, gholami2015flow1D,vidal2017convective} (supplemental Movie 100). In the opposite case of phosphodiestrase accumulation, pillars can act as wave sink (supplemental Movie 200). However, we should mention that, to prevent adsorption of chemicals to the PDMS substrate, we repeated our experiments with BSA (Bovine Serum Albumin) treated substrates and still observed concentric waves and periodic Voronoi domains around the pillars.  

In our system, we observed periodic domains within a range of caffeine concentration between 1-5 mM. Caffeine, which is a highly specific inhibitor of cAMP relay \cite{brenner1984caffeine,jaiswal2012regulation,steinboek1995spatial}, reduces both the cAMP production rate and wave frequency in a dose-dependent manner. We did not observe ordered aggregation territories without caffeine. Our experiments and earlier experiments by C. Weijer et.al.~\cite{siegert1989digital} with caffeine have shown that the aggregation territories are much larger in the presence of the caffeine compared to the controlled experiments. This means that the number of firing centers is decreased, which is also consistent with our numerical simulations in the presence of caffeine (higher values of $K_m$). We believe that in our experiments, a lower number of firing centers is crucial for circular waves emitted from pillars to successfully take over and dominate the system dynamics. We also observed that in the presence of obstacles, caffeine drastically diminished the number of spirals appearing in the system. Since spirals have a higher frequency than target patterns, they take over and dominate the dynamics once they appear (supplemental Movie 300). In other excitable systems, such as the BZ reaction it has been shown that spirals can be pinned to obstacles with lower excitability \cite{steinbock1992chemical}. Numerically it has been shown that the spiral tip interacts with the (no flux) obstacle boundary through attraction and repulsion \cite{munuzuri1998attraction}. We observe neither attraction nor repulsion of the spiral cores to the obstacles in both the experimental setup and the numerical simulations. We believe this crucial difference is given by the lack of meandering of the spiral tip in {\it D.d}.  Interestingly, it has been shown that in the presence of caffeine spiral core size increases in a dose dependent manner~\cite{steinboek1995spatial}.  In those occasions when a spiral did appear next to a pillar, it got pinned to the pillar and remained rotating around it (supplemental Movie 1000).
   
Comparing our numerical simulations with modified $K_m$ to the unmodified parameters, the system showed less centers, due to higher minimum cluster size necessary to produce pacemakers; it also showed smaller frequency, in agreement with experimental measurements (supplemental Figure 2). The minimum amount of cAMP necessary to produce cell activation was diminished, allowing for easier wave relay. Thus the transition boundary between "source" and "ignored" obstacles occurs at lower values of cAMP accumulation around the obstacles. For this reason, the presence of caffeine is necessary in our experiments to trigger formation of  concentric waves around the pillars.

From a general perspective, {\it D.d.} presents itself as a very suitable model system to further elucidate the mechanism of wave generation and synchronization in the presence of the external obstacles and investigations in this direction are underway in our lab.

\matmethods{
	\subsection*{Experimental Methods}
	The {\it D.d.} cells (strain AX2-214) were grown at 22$^{\circ}$C in HL5 medium, harvested in the exponential growth phase, and starved for four hours in 10 mL phosphate buffer supplemented with 2 mM caffeine in a shaking suspension~\cite{warren1976genetic}. After 4 hours of starvation, they were centrifuged and diluted to a density of $2 \times 10^6$cells/ml in fresh phosphate buffer containing 2 mM caffeine. The cell solution was transfered to a modified Petri dish with a plasma-treated polydimethylsiloxane (PDMS) substrate~\cite{whitesides2001soft}. The PDMS has a periodic array of macro-pillars characterized by pillar dimensions and spacing (see supplemental Fig. 1). If not stated otherwise, pillars of 1 mm diameter and height of 3 mm are arranged on square, triangular or hexagonal lattices with lattice size of 5 mm. cAMP wave patterns are indirectly visualized by dark-field microscopy. Since the change in scattered light reflects the cell shape change associated with cAMP, we indirectly observe the propagating waves of cAMP~\cite{alcantara1974signal,gross1976signal,devreotes1983quantitative,tomchik1981adenosine}.
	To allow better visualization, the frames in supplementary movie 2 are subtracted every 1 min and band-passed filtered to reduce the spatial noise. These processed images (supplemental movie 3) are used to calculate the phase at each pixel using the Hilbert transform (see supplemental movie 4).
\subsection*{Numerical Methods}
	The reaction-diffusion equations used for modeling this system are
\begin{subequations}
	\begin{align}
	k_1^{-1}\partial_t\rho_i&=-f_1(\gamma(x_i,y_i))\rho_i+f_2(\gamma(x_i,y_i))(1-\rho_i),\\
	\partial_t\beta_i&=s\Phi(\rho_i,\gamma(x_i,y_i))-(k_i+k_t)\beta_i ,\\ 
	\partial_t\gamma&=D\nabla^2\gamma-k_e\gamma + \sum_i^N H(i,x,y) k_t\beta_i/h,
	\end{align}
	\label{e:MartielGoldbeter_3Comp}
\end{subequations}
with
\begin{equation*}
\begin{array}{rlrl}
f_1(\gamma)&=\dfrac{1+\kappa\gamma}{1+\gamma},  &f_2(\gamma)&=\dfrac{\mathcal{L}_1+\kappa\mathcal{L}_2c\gamma}{1+c\gamma}, \\ \Phi(\rho,\gamma)&=\dfrac{\lambda_1+Y^2}{\lambda_2+Y^2}, 
&Y(\gamma,\rho)&=\dfrac{\rho\gamma}{1+\gamma},
\end{array}
\end{equation*}
and $s=q\sigma\alpha/(1+\alpha)$, where $\gamma(x,y)$ and $\beta_i$ are the amount of extracellular and intracellular cAMP respectively. $\rho_i$ corresponds to the percentage of active cAMP receptors on the cell surface and acts effectively as the slow variable that gives the system its excitable capabilities. $k_e$ corresponds to the extracellular phosphodiesterase and $s$ controls the amount of cAMP produced inside the cells. $H(i,x,y)$ is an index variable with values $1$ if the $i-th$ cells is located in $(x,y)$ and $0$ if it is not.
This produces that on the grid points containing amoebas the cAMP is transported from the intra- to the extracellular media, while on the empty cell spaces the wave is degraded by the unbounded phosphodiesterase.  
 The systems was simulated using discrete differences, a 5 point laplacian, and a time adaptive Runge-Kutta scheme. The used parameters were $\sigma=0.55$ min\textsuperscript{-1}, $k_1=0.09$ min\textsuperscript{-1}, $\kappa=18.5$,  $\mathcal{L}_1=10$, $\mathcal{L}_2=0.005$, $c=10$, $q=4000$, $\alpha=3$, $\lambda_1=10^{-4}$, $\lambda_2=0.2575$, $k_i=1.7$ min\textsuperscript{-1}, $k_t=0.9$ min\textsuperscript{-1}, $D=0.024$ mm\textsuperscript{2}/min, and $h=5$. With these parameters the system is in an oscillatory state, meaning that a limit cycle exists  along which a cell moves when it is isolated. If a cluster of cells of big enough size exists, it acts as a pacemaker producing trigger (chemical) waves that are relayed by the system, as long as a minimum percentage (measured as surface coverage) of cells exists. The dynamical details of the generation and relay of waves in this model will be presented elsewhere.
 Emulating the experimental observations, after some simulation time, when the waves have been established, we allowed the cells to be chemotactically competent. The movement rules were as follows. If a cell detects a cAMP gradient bigger than a threshold ($\nabla \gamma >g_{th}$) and it is in the refractory state ($\rho>\rho_{th}$), it moves against the gradient with a velocity $v_c$ as long as the two previous conditions continue to be fulfilled. If the new position falls in a different grid space, the movement occurs only if the new grid point does not already contain a cell. The parameters used were $g_{th}=1.0$, $\rho_{th}=0.6$, $v_c=200$ $\mu$m/min, and we allowed cell movement after $t=50$ min.     
}
\showmatmethods{} 

\acknow{We are grateful to E. Bodenschatz, V. Zykov, O. Steinbock, F. Mohammad-Rafiee, N. Visatemongkolchai, L. Turco, and I. Guido for fruitful discussions. T. E. acknowledges Deutsche Forschungsgemeinschaft (DFG), project Nr. GH 184/1-1. E.V.H. thanks the Deutsche Akademische Austauschdienst (DAAD), Research Grants - Doctoral Programs in Germany. A.G. acknowledges the MaxSynBio Consortium which is jointly funded by the Federal Ministry of Education and Research of Germany and the Max Planck Society.
}

\showacknow{} 


\bibliography{bibliography}

\newpage
\subsection*{Supporting Information (SI)}
%
%
\subsubsection*{Drift velocity of the annihilation point} 

We draw vertical lines starting at the annihilation point $B$ until the intersection with the wave $D$, and similarly we draw a line up from the later annihilation point $C$ up until the intersection $A$. The periods of the waves initiating from left and right pillars are $T_1$ and $T_2$, respectively (see Fig.~\ref{Drift}).    
Since $\overline{AC}$ and $\overline{BD}$ are parallel, the quadrilateral $ABCD$ is a scalene trapezoid with bases $\overline{AC}=T_1$ and $\overline{BD}=T_2$. The basal angles correspond with the wave propagation velocities such that $\angle BAC=\alpha$ and $\angle ACD=\beta$ with $\tan{\alpha}=v_1, \tan{\beta}=v_2$. 

Using the law of sines we calculate the other two sides of the trapezoid and arrive at 
$$\overline{AB}=(T_1-T_2)\dfrac{\sin{\beta}}{\sin{(\alpha+\beta)}}, \hspace{20 pt}
\overline{CD}=(T_1-T_2)\dfrac{\sin{\alpha}}{\sin{(\alpha+\beta)}}.$$
We now look at the angle $\angle CBD=\gamma$ such that $\tan{\gamma}=v_d$ is the drift velocity of the annihilation point. Applying law of sines to the triangle $\triangle CBD$ we get
$$ \dfrac{\sin{(\angle BCD)}}{\overline{BD}}=\dfrac{\sin{\gamma}}{\overline{CD}}. $$ 
Substituting with the values already calculated we get
$$ \dfrac{\sin{(\beta-\gamma)}}{T_2}=\dfrac{\sin{\gamma}}{(T_1-T_2)}\cdot \dfrac{\sin{(\alpha+\beta)}}{\sin{\alpha}}, $$
expanding and simplifying we obtain 

$$(T_1-T_2)\left(\dfrac{1}{v_d}-\dfrac{1}{v_2} \right)=T_2\left(\dfrac{1}{v_1}+\dfrac{1}{v_2} \right),$$
finally we arrive to 
\begin{equation}
v_d=\dfrac{T_1-T_2}{T_1/v_2 + T_2/v_1},
\end{equation}
which in the limit case of $v_1=v_2=v_w$ reduces to
\begin{equation}
v_d=v_w \dfrac{T_1-T_2}{T_1 + T_2}.
\end{equation}
\subsubsection*{SI Figures}
\setcounter{figure}{0} 
  \renewcommand{\figurename}{Fig. S} 
\begin{figure}[htbp]
\begin{center}
	\includegraphics[width=0.284\columnwidth]{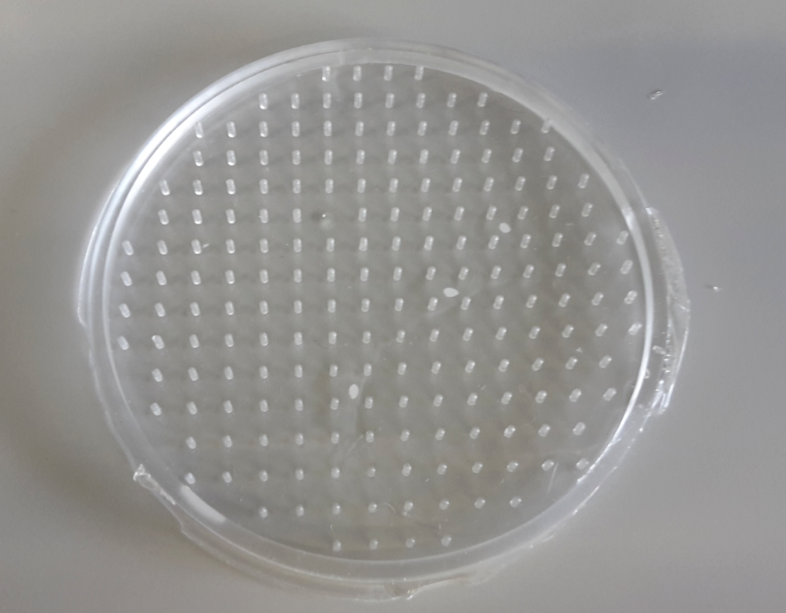}
	\includegraphics[width=0.7\columnwidth]{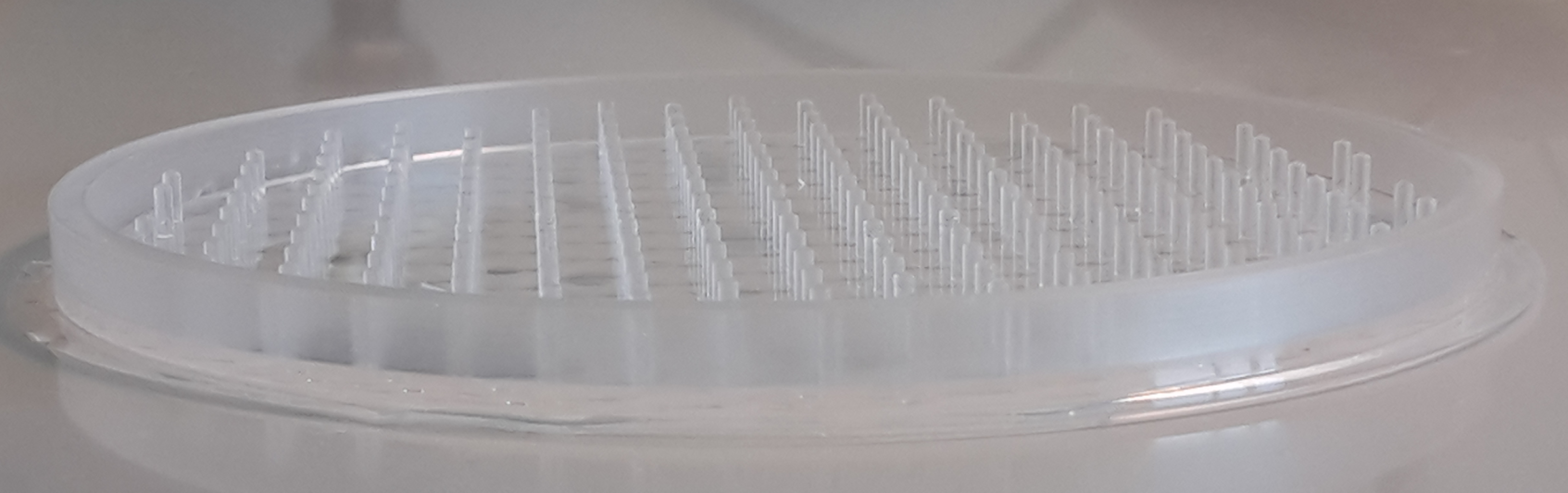}
	\caption{Top view and side view of the PDMS block with a quadratic arrangement of the pillars that fits to a normal 100 mm plastic Petri dish. The pillars are 1 mm thick, 3 mm high and are spaced 5 mm away from each other.}
	\label{setup}
\end{center}
\end{figure} 
\begin{figure}[htbp]
	\includegraphics[width=0.7\columnwidth]{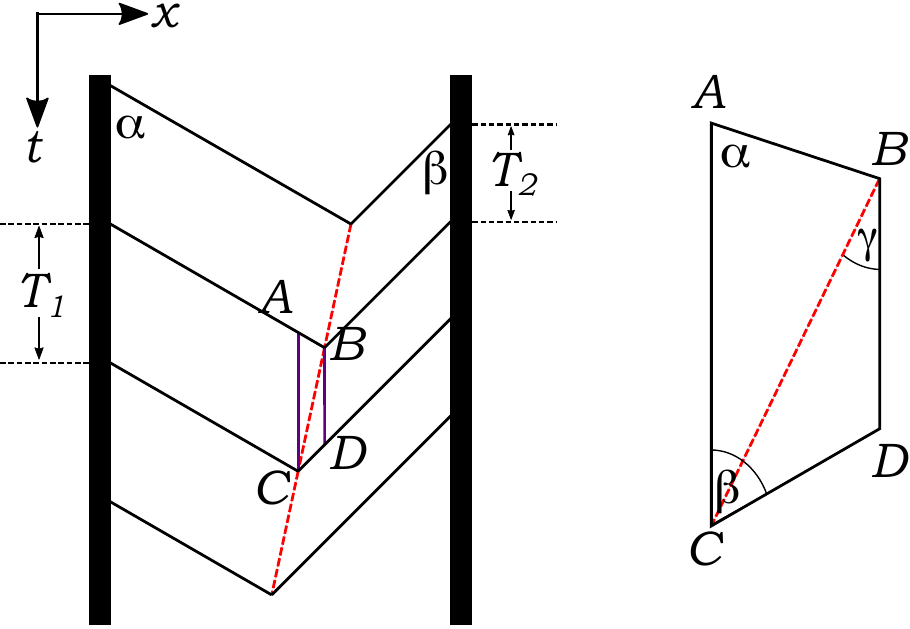}
	\caption{Schematic representation in a space-time plot of two waves coming from pillars. Thick lines represent pillars and the red dashed line marks the position of the annihilation point. Waves are emitted from the left pillar with a period $T_1$ and with $T_2$ from the right pillar. These values correspond to the vertical spacing between waves in this representation.}
	\label{Drift}
\end{figure} 
\begin{figure}[htbp]
	\includegraphics[width=0.7\columnwidth]{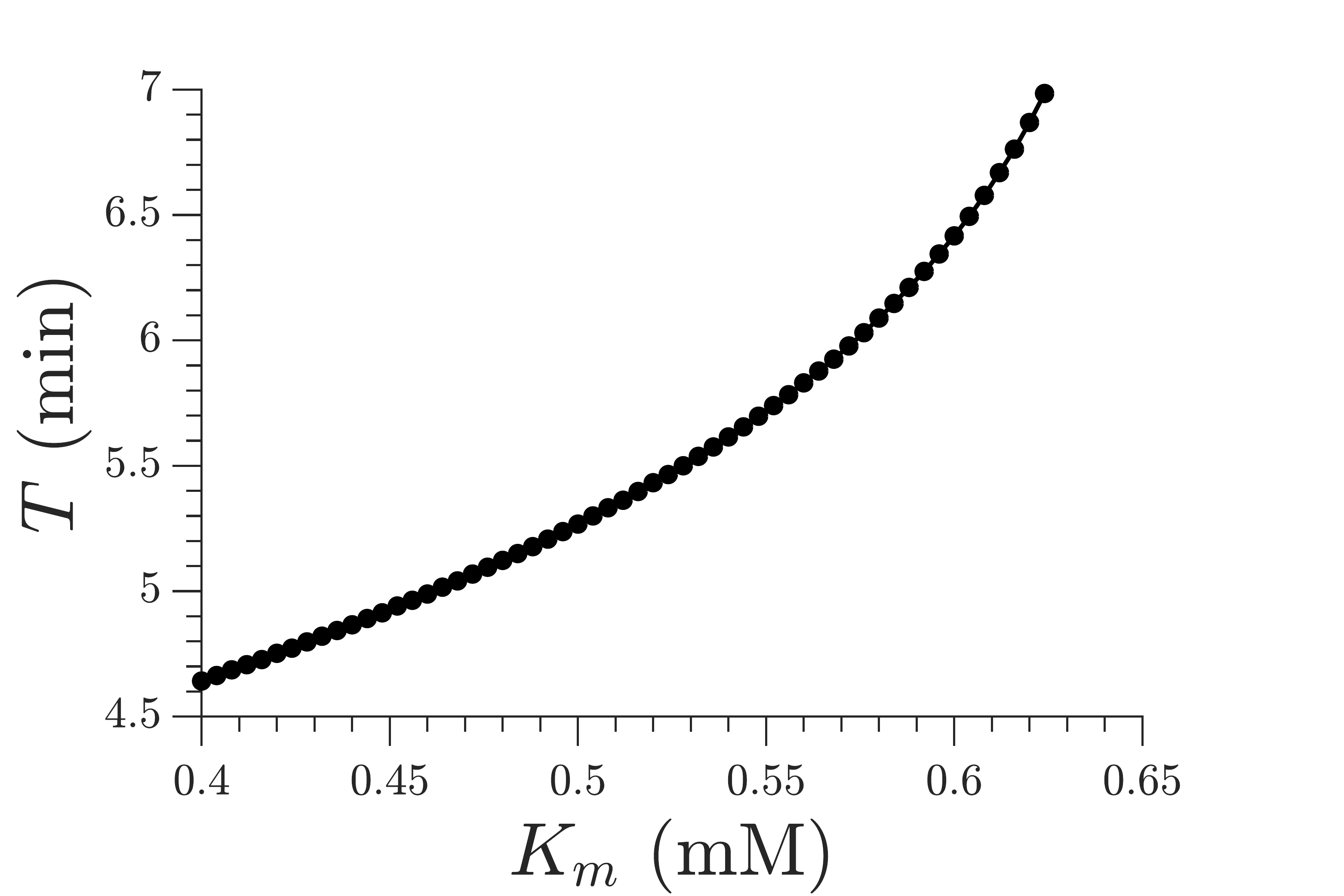}
	\caption{The wave period increases at higher $K_m$ values (higher caffeine concentrations).}
	\label{Period-Km}
\end{figure}

%
%
%
%
%
%
%

\end{document}